\documentclass[11pt,twoside]{article}

\usepackage{asp2006}
\usepackage{epsf}
\usepackage{lscape}
\usepackage{graphicx}

\begin{document}
\title{The Tremaine-Weinberg method for Pattern Speeds using H$\alpha$ emission from Ionized Gas}   
\author{John E. Beckman$^{(1,2)}$; Kambiz Fathi$^{(1,3)}$; N\'uria Pi\~nol$^{(1,4)}$; Silvia Toonen$^{(1,5)}$; Olivier Hernandez$^{(6)}$; Claude Carignan$^{(6)}$}   
\affil{	(1) Instituto de Astrof\'\i sica de Canarias, La Laguna, Spain\\
	(2) Consejo Superior de Investigaciones Científicas, Spain\\
	(3) Stockholm Observatory, Stockholm University, Sweden\\
	(4) Departamento de F\'\i sica - CIOyN, Universidad de Murcia, Spain\\
	(5) Leiden Observatory, Leiden University, The Netherlands\\
	(6) LAE, Dept. Physics, Universite de Montr\'eal, Montr\'eal, QC Canada}    

\begin{abstract} 
The Fabry-Perot interferometer FaNTOmM was used at the 3.6m Canada France Hawaii Telescope and the 1.6m Mont M\'egantic Telescope to obtain data cubes in H$\alpha$ of 9 nearby spiral galaxies from which maps in integrated intensity, velocity, and velocity dispersion were derived. We then applied the Tremaine-Weinberg method, in which the pattern speed can be deduced from its velocity field, by finding the integrated value of the mean  velocity along a slit parallel to the major axis weighted by the intensity and divided by the weighted mean distance of  the velocity points from the tangent point measured along the slit. The measured variables can be used either to make separate calculations of the pattern speed and derive a mean, or in a plot of one against the other for all the points on all slits, from which a best fit value can be derived. Linear fits were found for all the galaxies in the sample. For two galaxies a clearly separate inner pattern speed with a higher value, was also identified and measured. \\[-2cm]
\end{abstract}


\section{Overview}   
The theory of resonant structure in disk galaxies in its original linear form by Lindblad (1963) or in its more complete form by Lin \& Shu (1966) implies that a density wave pattern in the stellar disk acts on the rotating gas to form stars continually along spiral shock lines associated with the arms or virtually straight shocks associated with the major bar. One corollary is that over significant ranges of galactocentric radius the angular velocity of the density wave pattern may well be virtually invariant. These ranges are associated with bars and spiral arms, and their respective angular velocities are termed pattern speeds, $\Omega_p$. Observational derivation of $\Omega_p$, together with the conventional rotation curve, allows us to characterise the main dynamical parameters of the disk. Tremaine \& Weinberg (1984) (TW) suggested a virtually model independent purely kinematic way to derive pattern speeds. To make basic use of the method we need to take a spectrum along a slit parallel to the major axis, and calculate $\Omega_p$ using the velocity and luminosity in the component used to measure the spectrum. A superior method uses a two dimensional velocity field to simulate a large number of slits, enhancing the S/N and better exploring the regime of multiple pattern speeds. Here we have used Fabry-Perot cubes in H$\alpha$ emission which provide complete velocity fields in two dimensions and also an H$\alpha$ surface brightness map, an ideal data set for the TW method.

\parbox{0.52\textwidth}{
  \includegraphics[width=.51\textwidth]{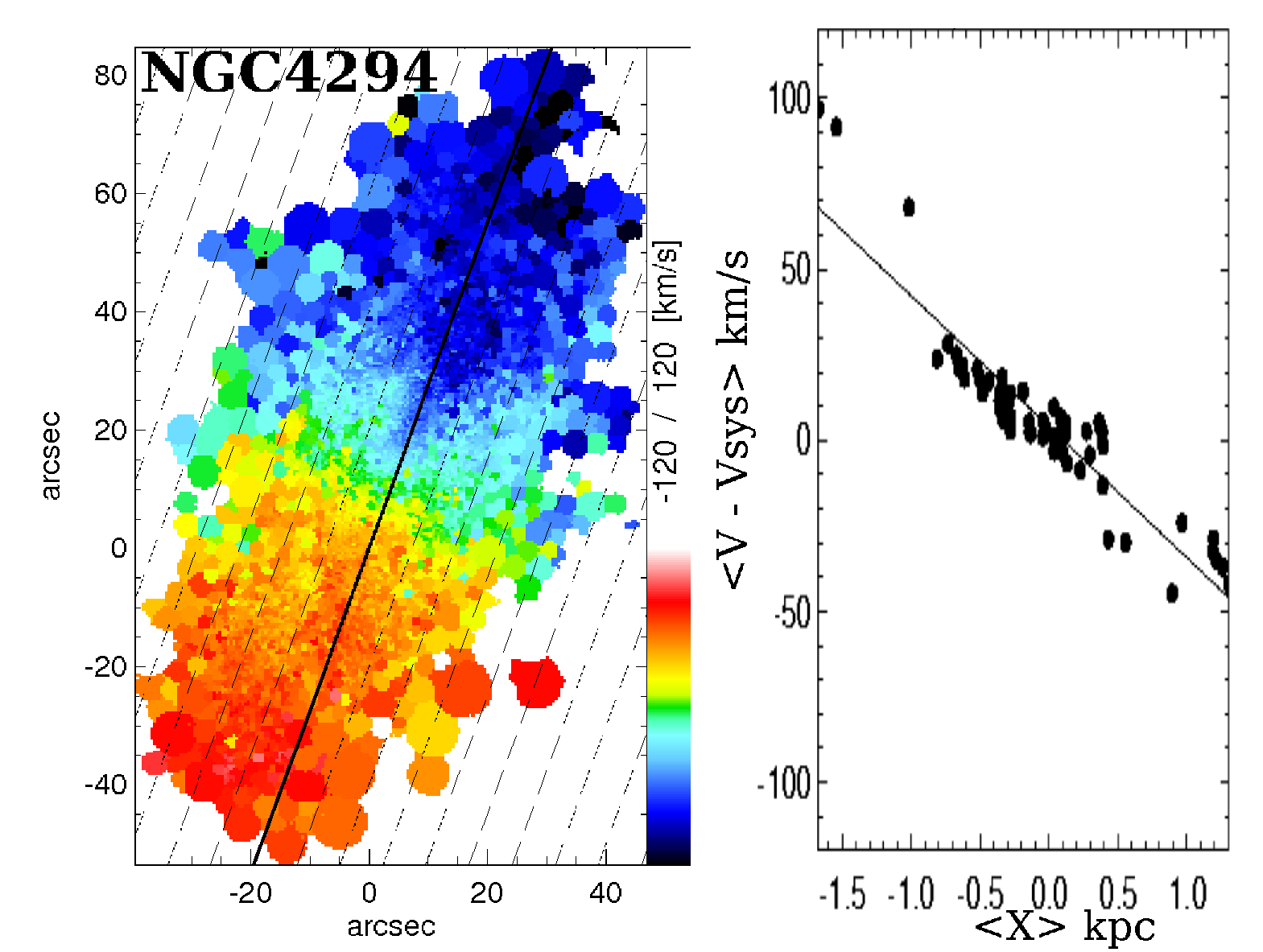}}
\parbox{0.20\textwidth}{
\begin{tabular}{ccc}\hline
NGC&  $\Omega_p$ main	     &   $\Omega_p$ inner \\	   
   &  $\overline{km/s/kpc}$  &   $\overline{km/s/kpc}$ \\\hline       
3049   &  32   &   68 \\
4294   &  42   &   -- \\
4519   &  13   &   -- \\
5371   &  16   &   -- \\
5921   &  13   &   -- \\
5964   &  23   &   -- \\
6946   &  22   &   47 \\
7479   &  18   &   -- \\
7741   &  33   &   -- \\\hline 
\end{tabular}}\\
{\em Example of extracting the velocities from the H$\alpha$ data (left), and initial estimates for the pattern speed values for the entire sample (right).}\\
There are two caveats here: Firstly the method requires that the observed component obeys the continuity equation. In principle this may not be the case for H$\alpha$-emitting gas, subject to large scale shocks and smaller scale expansions around major OB associations. Our main claim here is that "the proof of the pudding is in the eating", as shown by the excellent linear fits and the clear separations of the pattern speeds of known inner bars (e.g., Fathi et al. 2007). Clear support comes from Hernandez et al. (2005) who found that this technique gave a triple pattern speed for NGC\,4321.  Secondly we need a valid position angle for the disk line of nodes. Here we used both kinematic and morphological symmetry to derive our position angles. Both of these points need very careful treatment in practical pattern speed derivation and will be treated fully in a a forthcoming article. In the table above, we give the full set of all the $\Omega_p$ values.\\[-5mm]

\section{Summary}    
We have shown that an intensity-velocity map in H$\alpha$ of the type obtained using a Fabry-Perot two-dimensional spectrograph, can be used to derive the gas kinematics of disk galaxies with coverage and precision sufficient to derive pattern speeds for those where star formation is well spread across the galaxy. Although this technique is not new, its precision, and the number of galaxies analysed here give our results considerable value. These pattern speeds complement the values for 7 galaxies, of earlier types than those selected here, by Corsini et al. (these proceedings), and should allow us to initiate interesting dynamical analysis.

\acknowledgements 
This work was supported by projects AYA2004-08251-C02-01 and AYA2007-47625-C02-01 of the Spanish Ministry of Education and Sciences and P3/86 of the IAC. It is a pleasure to thank the organisers for a stimulating conference.

\end{document}